\documentclass[pre,onecolumn,draft,showpacs,preprint,amsmath,
amssymb,floatfix]{revtex4}
\usepackage{bm}
\begin{document}
\flushbottom
\input{psfig.tex}
\title{Synchronization Based Approach for Estimating
All Model Parameters of Chaotic Systems}
\author{Rahul Konnur}
\affiliation{Tata Research Development and Design Centre, \\
54B, Hadapsar Industrial Estate, Pune. 411 013. India.} 
\date{\today}
\begin{abstract}
The problem of dynamic estimation of all parameters of a model
representing chaotic and hyperchaotic systems using information
from a scalar measured output is solved. The variational calculus
based method is robust in the presence of noise, enables online
estimation of the parameters and is also able to rapidly track
changes in operating parameters of the experimental system.
The method is demonstrated using the Lorenz, Rossler chaos and
hyperchaos models.
Its possible application in decoding communications using chaos is 
discussed.
\end{abstract}
\maketitle

Synchronization of uni-directionally coupled chaotic systems has been
a subject of great interest for over a decade
\cite{ref1,ref2,ref3,ref4}. 
The interest in understanding the synchronization characteristics of 
chaotic systems stems from its potential applications in a variety of
areas, for e.g. in optics, communications and time series analysis of
chaotic systems. 
An important issue in time series analysis of chaotic systems is the
estimation of all parameters using information from a
scalar measured output. 
This information,
can in turn be used to estimate all the unmeasured system states,
provided the model is known.

In general there are three key issues
in parameter estimation of dynamical systems. Firstly, the method has
to be robust in the presence of noise. Secondly, the method must
allow estimation of all parameters using 
any conveniently measurable output from the system. Thirdly, it must
be able to rapidly track changes in the operating parameters of the experimental system. 

Current parameter estimation techniques can be broadly classified
as online and off--line strategies. 
The online, e.g. adaptive control approach \cite{ref5}, though
simple to implement, has been demonstrated to be unsuitable
for estimation of multiple parameters (e.g. for the Rossler system).
In contrast, off--line, e.g.
autosynchronization \cite{ref6} and error minimization
\cite{ref7,ref8} schemes have been
demonstrated to be able to estimate all parameters.
The former is a geometric approach where the optimal
vector fields governing temporal evolution of the parameters are
obtained using a linearization based numerical procedure.
Using the error minimization approach, Goodwin et al. have shown that
all parameters can be estimated by using a scalar measurable output
(not necessarily corresponding to one of the state space variables) 
of chaotic and hyperchaotic systems \cite{ref8}.

For systems with multiple parameters, the least squares error function
possesses several minima \cite{ref8}. This can lead to
an erroneous estimate of parameters owing to convergence at one of
the local minima.
Moreover, in practice, it may not be always possible to
possess information of the time at which
parameter changes in the experimental system occur.
For online schemes, a lack of this information results in erroneous
estimates of the determined parameters.
Off--line schemes cannot be used in this situation, since there is no
way to incorporate the effect
of parameter changes in the least squares minimization
method traditionally employed for parameter estimation. 

In this Report, the least squares approach is used to develop
a general and robust method for deriving
the dynamical system governing the evolution of all parameters of
a chaotic system. The technique is demonstrated using simulated
experimental data from the Lorenz, Rossler chaos and Rossler
hyperchaos systems.
The advantages of the method
are its ability to: (a) estimate all parameters in an online setting,
(b) respond to unknown changes in parameters of the
experimental system, and 
(c) converge to the `true' estimates of the parameters. 
Each of these is a distinct improvement over
capabilities of the existing methods.

We begin by briefly describing the set up of the parameter estimation problem. Let
\begin{equation}
{\bf{\dot x}}={\bf{f}}({\bf{x,p}})
\end{equation}
\noindent represent the experimental dynamical system
with state variables ${\bf{x}} \in {\bf{R}}^n$, whose parameters 
${\bf{p}} \in {\bf{R}}^m$ are to be estimated.
 The overdot indicates differentiation with respect to time
$t$. The only information available from this experimental system is
(i) the functional form of the model and (ii) a scalar time series
given by an observable $s({\bf{x}})$. The model is given by the 
following equation
\begin{equation}
{\bf{\dot y}}={\bf{g}}({\bf{y,q}})
\end{equation}
where ${\bf{y}} \in {\bf{R}}^n$ and ${\bf{q}} \in {\bf{R}}^m$, and the
functional form of ${\bf{g}}$ is identical to that of ${\bf{f}}$  in (1).
{\em We assume that there exists an uni-directional coupling scheme 
using the available scalar output of the experimental system (1) which
enables asymptotic synchronization of the model system (2) with the 
experimental system (1), i.e. ${\bf{y}} \rightarrow {\bf{x}}$ as
$t \rightarrow \infty$, if ${\bf{q}} = {\bf{p}}$.}
The coupling can either be a drive/response coupling scheme, 
e.g. the Pecora-Carroll scheme \cite{ref1} or a feedback coupling
scheme \cite{ref9}. The conditions which enable synchronization of
the model system (2) with the experimental system (1) are well known.
In most cases, synchronization is guaranteed if all conditional
Lyapunov exponents of the error system 
${\bf{\dot e}} = {\bf{\dot x}} - {\bf{\dot y}}$ constructed using 
Eqs.(1) and (2) are negative \cite{ref10}. 
Techniques are now available which enable design of an uni-directional
scheme which guarantees synchronization \cite{ref11}. 

For the sake of conciseness, we only consider feedback coupling in
this report. The general representation of this scheme is given by
\begin{equation}
{\bf{\dot y}}={\bf{g}}({\bf{y,q}}) - {\bf{BK}}^T (s({\bf{y}})-s({\bf{x}}))
\end{equation}
\noindent where ${\bf{B}}$  is a constant vector and ${\bf{K}}$ is the
gain vector. The scheme (3) has been previously used for achieving
identical synchronization in a hyperchaotic system \cite{ref9}. 

We now develop the main theme of this Report which can be defined as:
``Develop a formalism for constructing a system of differential
equations governing the evolution of the model system parameters
${\bf{q}}$ such that $({\bf{y,q}}) \rightarrow ({\bf{x,p}})$  as 
$t \rightarrow \infty$."
Our objective will be to design a parameter evolution scheme that 
asymptotically drives the measured error 
$s({\bf {y}}) - s({\bf {x}})$ to zero, and thereby yields 
$({\bf{y,q}}) \rightarrow ({\bf{x,p}})$ as $t \rightarrow \infty$. 
The starting point is the following minimization problem
\begin{equation}
G({\bf{q}})=\min[(s({\bf{y}})-s({\bf{x}}))^2]
\end{equation}
We note that an inability to correctly estimate the
initial conditions of the state variables and/or parameters 
could result in large errors during the initial stage of evolution of
the model system. Our goal, reflected in the choice of the cost
function (4), is to force the model output $s({\bf{y}})$ to 
asymptotically synchronize with the experimental output $s({\bf {x}})$.
The minimization problem (4) is studied as the following equivalent 
system of differential equations
\begin{equation}
\dot q_j = -\frac{\partial{G}}{\partial{q_j}} = 
-2(s({\bf{y}})-s({\bf{x}}))
\frac{\partial s({\bf{y}})}{\partial{q_j}}, \ \ \ j=1,\cdots, m
\end{equation}
The equilibrium state
of the system (5) is typically attained when the synchronization condition
is satisfied, i.e. when $s({{\bf y}})=s({{\bf x}})$. This ensures
that the
parameter estimates attain their true values. In contrast,
convergence to the true parameters is not guaranteed in the error
minimization approach since the least squares cost function has
several local minima \cite{ref8}.
A knowledge
of the variational derivatives ${\partial {y_i}}/{\partial {q_j}}$
for $i=1,\cdots, n$ and $j=1,\cdots, m$ is needed for solving this
system of equations. Since the functional form of the model is known,
these derivatives are given by (using Eq. (3))
\begin{eqnarray}
\frac{d}{dt} \left( \frac{\partial y_i}{\partial q_j} \right)& = &
\sum_{k=1}^{n} \frac{\partial g_i}{\partial y_k} \frac{\partial y_k}
{\partial q_j} + \frac{\partial g_i}{\partial q_j} - 
{{\bf {B}} {\bf{K}}}^T \sum_{k=1}^{n} \frac{\partial s({\bf{y}})}
{\partial q_j} \nonumber \\
& & i=1,\cdots, n; \ \ j=1,\cdots,m
\end{eqnarray}
Formulation of Eqs. (4)--(5) and using Eq. (6) to solve Eq. (5) are
the key steps in the proposed procedure for estimating all parameters
of a chaotic/hyperchaotic system. The method consists of solving
(i) the experimental system (1) 
(when real experimental data is not available),
(ii) the model system (3), (iii) the equations
\begin{equation}
\dot q_j = -\epsilon_j\frac{\partial{G}}{\partial{q_j}} = 
-2\epsilon_j(s({\bf{y}})-s({\bf{x}}))
\frac{\partial s({\bf{y}})}{\partial{q_j}}, \ \ \ j=1,\cdots, m
\end{equation}
governing evolution of the parameters, and (iv) the equations
corresponding to the evolution of the variational derivatives
Eq. (6). 
The vector of additional parameters ${\bf{\epsilon}}$ is needed for 
guaranteeing stability of the overall system and it also controls
the rate of synchronization. When the actual
scalar measured output from an experiment is available, an extended
system comprising of $n+m+nm$ equations needs to be solved in order
to estimate $m$ parameters of a $n$ dimensional system. The condition
for convergence of the procedure is that the real part of the
eigenvalues of the Jacobian matrix or the conditional Lyapunov
exponents of the extended system formed using Eqs. (3) and (7) 
are all less than zero \cite{ref6}.

Our first example is the Lorenz system. 
We demonstrate the method when the scalar observable is $x_2$.
The simulated experimental and model systems are respectively given
by the following equations 
\begin{equation}
\begin{array}{llllll}
\dot x_1&=&p_1(x_2-x_1)\  & \dot y_1&=&q_1(y_2-y_1) \\
\dot x_2&=&p_2x_1-x_2-x_1 x_3 \ & \dot y_2&=&q_2y_1-y_2-y_1 y_3 \\
& & & & & -k(y_2-x_2) \\
\dot x_3&=&x_1x_2-p_3x_3 \ & \dot y_3&=&y_1y_2-q_3y_3 \\
\end{array}
\end{equation}
with $p_1=10$, $p_2=28$ and $p_3=8/3$.
The six equations (8), together with four equations (Eqs. (6) and (7))
for estimation of
each of the three parameters results in a set of eighteen equations,
governing the evolution of the (i) simulated experimental system, 
(ii) model
system, (iii) parameters and (iv) the variational derivatives. When
${{\bf p}}={{\bf q}}$, the feedback term corresponding to the 
product of the constant vector ${\bf{B}}=[0,1,0]^T$, suitable gain
${\bf {K}}=[0,k,0]^T$ and the output error $(y_2-x_2)$ guarantees 
synchronization of the model system with the experimental system.
We consider the situation where additive uniformly distributed 
random noise
in the range $[-0.5,0.5]$ is present in the measured output $x_2$.
Figure 1 shows the evolution of the parameters and the relative
estimation errors for the case where in
addition to noise, a step perturbation in parameters is imposed on
the simulated experimental system.
{\em It is important to note that this introduces an additional 
complexity since information about the imposed perturbation is not
available to the model}. 
The robustness of the method is demonstrated by the convergence to
the original parameters close to $t=125$; followed by a rapid, stable transition
and subsequent convergence into the vicinity of the new operating
parameters. All parameters could be successfully determined when
the measured output was the $x_1$ variable. However,
the method fails when the measured output is the $x_3$ variable, since
the Lyapunov exponents of Eq. (3) are not negative for any choice of
${\bf B}$ and ${\bf K}$.

Our next example is the Rossler system and we demonstrate
the method when the scalar observable is $x_2$. The simulated
experimental and model system equations are given by 
\begin{equation}
\begin{array}{llllll}
\dot x_1&=&-x_2-x_2)\  & \dot y_1&=&-y_2-y_3 \\
\dot x_2&=&x_1+p_1x_2 \ & \dot y_2&=&y_1+q_1y_2 \\
& & & & & -k(y_2-x_2) \\
\dot x_3&=&p_2+x_3(x_1-p_3) \ & \dot y_3&=&q_2+y_3(y_1-q_3) \\
\end{array}
\end{equation}
with $p_1=0.2$, $p_2=0.2$ and $p_3=9$.
The feedback parameter vectors were selected to be ${\bf{B}}=[0,1,0]^T$
and ${\bf {K}}=[0,k,0]^T$. Results for two different cases are shown
in Fig. 2. These correspond to the situation when (i) additive
uniformly distributed random noise in the range
$[-0.1,0.1]$ is present in the measured output [Fig. 2(a)] and
(ii) in addition to noise, 
the parameters of the simulated experimental system are increased by
$10\%$ at $t=500$ [Fig. 2(b)].
For both cases, the parameter estimates exhibit small-amplitude
fluctuations around the correct
value. 
The nature of evolution of errors in Fig. 2(b) indicates that the
step perturbation results in a rapid transition
of the parameter estimates into the vicinity of the new value. 
A feature of this result
is the direct dependence of the magnitude of fluctuations of 
individual parameters on the manner in which the parameter is related
to the measured output. 

The next example is a four parameter Rossler hyperchaos system.
The simulated experimental and model systems are
given by the following equations 
\begin{equation}
\begin{array}{llllll}
\dot x_1 & = & - x_2 - x_3 & \dot y_1 & = & - y_2 - y_3 \\
\dot x_2 & = & x_1 + p_1x_2 + x_4 & \dot y_2 &=& y_1+q_1y_2+y_4-w \\
\dot x_3 & = & p_2 + x_1 x_3 & \dot y_3 &=& q_2 + y_1 y_3 \\
\dot x_4 & = & -p_3x_3 + p_4x_4 & \dot y_4 &=& -q_3y_3 + q_4 y_4 -w \\
\end{array}
\end{equation}
with $p_1=0.25$, $p_2=3$, $p_3=0.5$ and $p_4=0.05$.
The measurable output is assumed to be $s({\bf{x}})=x_2+x_4$,
$w=k(s({\bf{y}})-s({\bf{x}}))$,
${\bf{B}}=[0,1,0,1]^T$ and ${\bf{K}}=[0,k,0,k]^T$. It is possible
to achieve identical synchronization for this system using 
uni-directional coupling and a suitable choice of ${{\bf K}}$. 
We study the ability of the method to track changes in operating
parameters of the simulated experimental system. It is assumed that
parameters
$p_1$ and $p_2$ of the simulated experimental system are increased by
$4\%$ at $t=500$. The scaled temporal evolution of all the 
estimated parameters $q_i,i=1,\cdots,4$ is shown in Fig. 3(a). 
Figure 3(b) shows the evolution of the estimated parameters for the
case when in addition to changes in operating parameters of the 
simulated experimental system, uniformly distributed random
noise in the
range  $[-0.005,0.005]$ is present in the experimental measured
output. For each case, it can be seen that the method allows a rapid
convergence to the new operating parameters.

Finally, we present a simple example to illustrate a possible application
of the proposed parameter estimation scheme in communications using
chaos. The problem relates to decoding
of an encoded message signal. It is assumed that the following 
information is known: (i) the chaotic system used to encode the
message, and (ii) the state variable used for encoding the message. 
For simplicity, the message is taken to be a sinusoidal function, 
and it is assumed that the encoding
is carried out in an additive manner. The objective is to decode the
noisy transmitted signal and retrieve the message. The first step in achieving this
objective is to estimate the parameters of the model system. Once this
has been accomplished, the message can be retrieved by subtracting the
computed model output from the transmitted signal. Figure 4 shows the
fair degree of comparison between the message signal and the decoded
signal estimated using the Lorenz and Rossler systems for the case 
when additive noise in the range $[-0.1,0.1]$ is also present in
the transmitted signal.

In conclusion, we have developed an analytical framework for the
robust design of dynamical systems that guarantees online
estimation of all model parameters of a given chaotic/hyperchaotic
system when certain conditions are satisfied. 
The key advantage of the proposed technique is that it requires information
of only a scalar measured output from the experimental system. 
A possible application in communications using chaos has been demonstrated. 
We have verified that the method
proposed here would be applicable to the more realistic situation
where only discrete-time measurements of the experimental output
are available \cite{ref8}. These results will be communicated in the future.

\acknowledgements{The author gratefully acknowledges funding and
support provided by the Tata Research Development and Design Centre
(TRDDC), Pune, India.}

\newpage
\begin{center}
{\bf List of Figures}
\end{center}
\noindent {\bf FIGURE 1}\ Temporal evolution of (a) all the 
three parameters 
of the Lorenz system, (b) fractional relative errors
for the case when additive noise is
present in the measured output and the parameters
of the simulated experimental system are changed to $p_1=11$,
$p_2=35$ and $p_3=3$ at $t=150$. The stability parameters are 
$k=25$, $\epsilon_1=1$, $\epsilon_2=15$ and $\epsilon_3=1$.
The straight lines in (b) correspond to an error of $\pm 5\%$. \\ \\
\noindent{\bf FIGURE 2}\ Temporal evolution of the estimation 
errors for all the 
parameters of the Rossler system (9) for the case when (a) only
additive noise is present in the measured output $x_2$,
(b) additive noise is present and each of the parameters of the 
simulated
experimental system is increased by $10\%$ at $t=500$. The
parameter $q_3$ has been scaled down 50 times for greater clarity of
representation. 
The stability parameters are  $k=20$, $\epsilon_1=0.15$, 
$\epsilon_2=0.2$ and $\epsilon_3=2$. The straight lines
correspond to an error of $\pm 5\%$. \\ \\
\noindent{\bf FIGURE 3}\ Temporal evolution of the four parameters
$q_1$, $q_2$,
$q_3$ and $q_4$ of the Rossler hyperchaos system (10) for the case
when (a) the parameters $p_1$ and $p_2$ are increased by $4\%$ at 
$t=500$, (b) additive noise is present in the 
measured output $x_2+x_4$, and the parameters $p_1$ and $p_2$ of the
simulated
experimental system are increased by $4\%$ at $t=300$.
For greater clarity of representation, the parameters have been
scaled in the following manner: $q_2$  and $q_3$  are scaled down
10 and $\frac{3}{4}$ times respectively, while the parameter $q_4$ 
is scaled up 4 times. The stability parameters are (a) $k=4$, 
$\epsilon_1=0.75$ and $\epsilon_2=\epsilon_3=\epsilon_4=0.005$; 
(b) $k=3.5$, $\epsilon_1=0.80$ 
and $\epsilon_2=\epsilon_3=\epsilon_4=0.002$. \\ \\
\noindent{\bf FIGURE 4}\ Comparison of the message signal 
$I=\sin(20\pi t)$ and the
decoded signals obtained using the $x_2$ variable of Lorenz and 
Rossler models as the measured output. These are represented by the 
continuous, dashed and dotted lines respectively.  
Decoding results in phase shift as well as in a reduction
of the amplitude. The amplitude of the message
signal has been scaled for greater clarity. The stability parameters
are the same as that given in Fig. 1 and Fig. 2.
\newpage
\begin{figure}[htbp]
\centerline{{\psfig{file=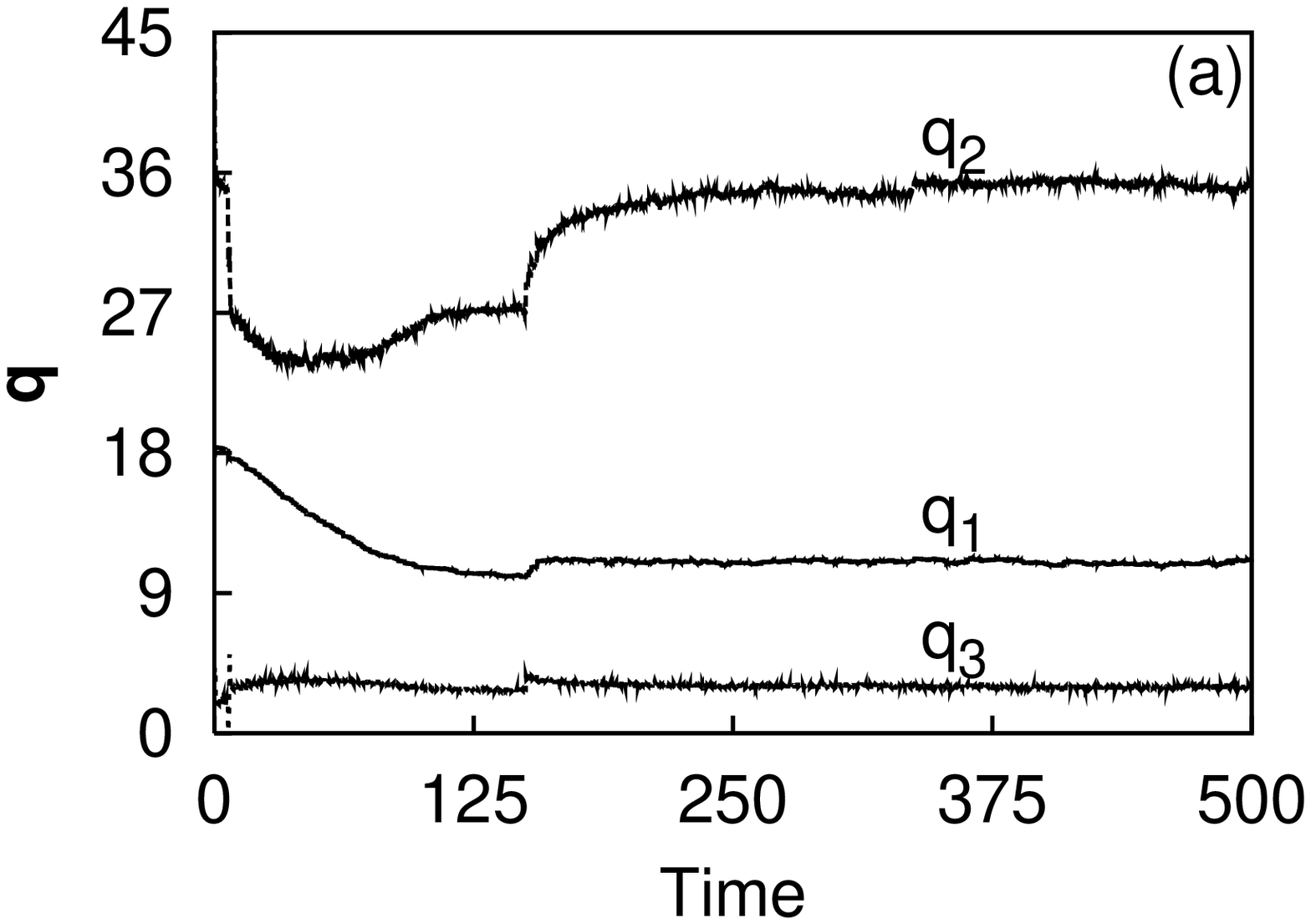,width=1.60in}}
{\psfig{file=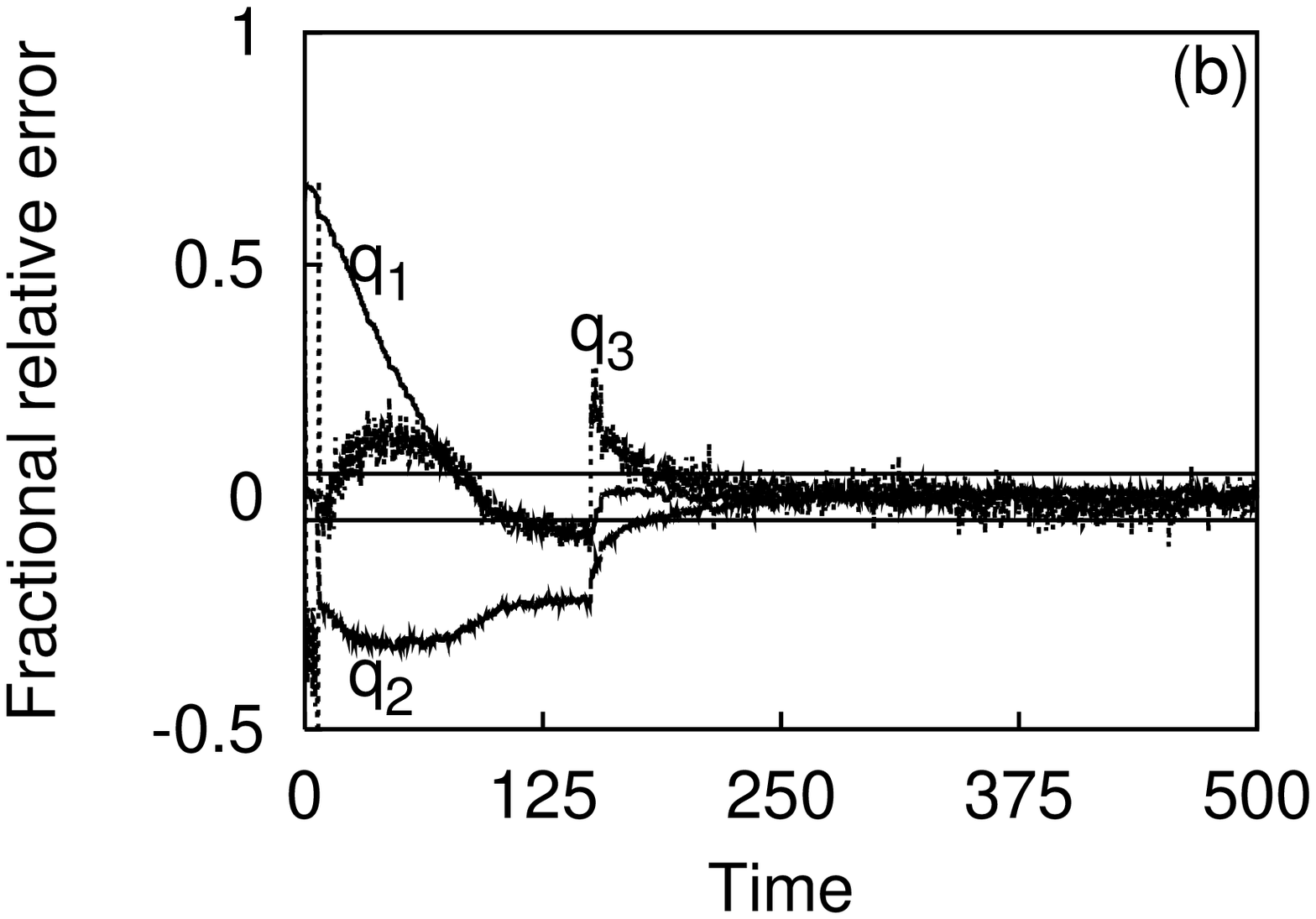,width=1.60in}}}
\caption{Temporal evolution of (a) all the three parameters 
of the Lorenz system, (b) fractional relative errors
for the case when additive noise is
present in the measured output and the parameters
of the simulated experimental system are changed to $p_1=11$,
$p_2=35$ and $p_3=3$ at $t=150$. The stability parameters are 
$k=25$, $\epsilon_1=1$, $\epsilon_2=15$ and $\epsilon_3=1$.
The straight lines in (b) correspond to an error of $\pm 5\%$.}
\end{figure}
\newpage
\begin{figure}[htbp]
\centerline{\psfig{file=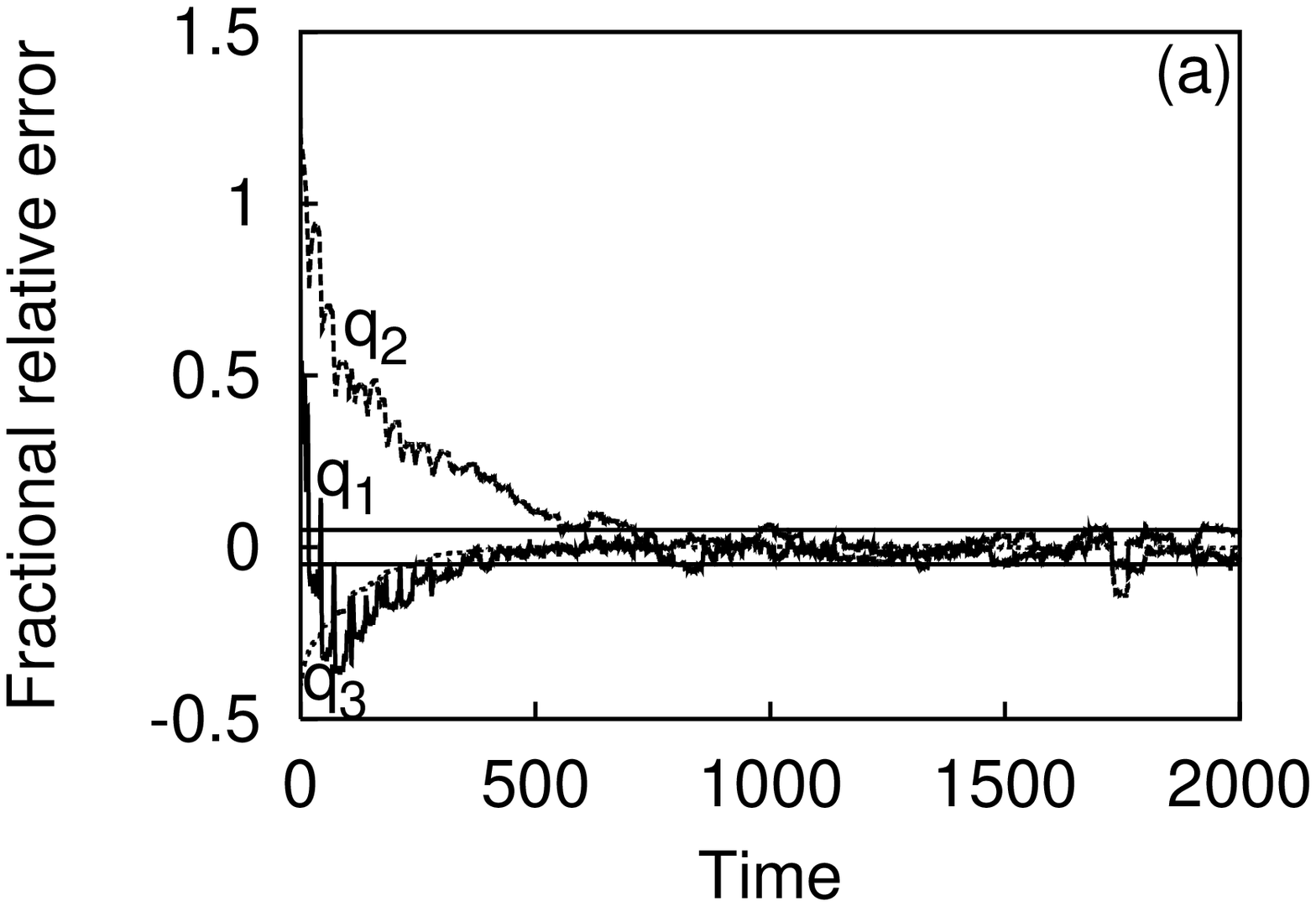,width=2.30in}}
\centerline{\psfig{file=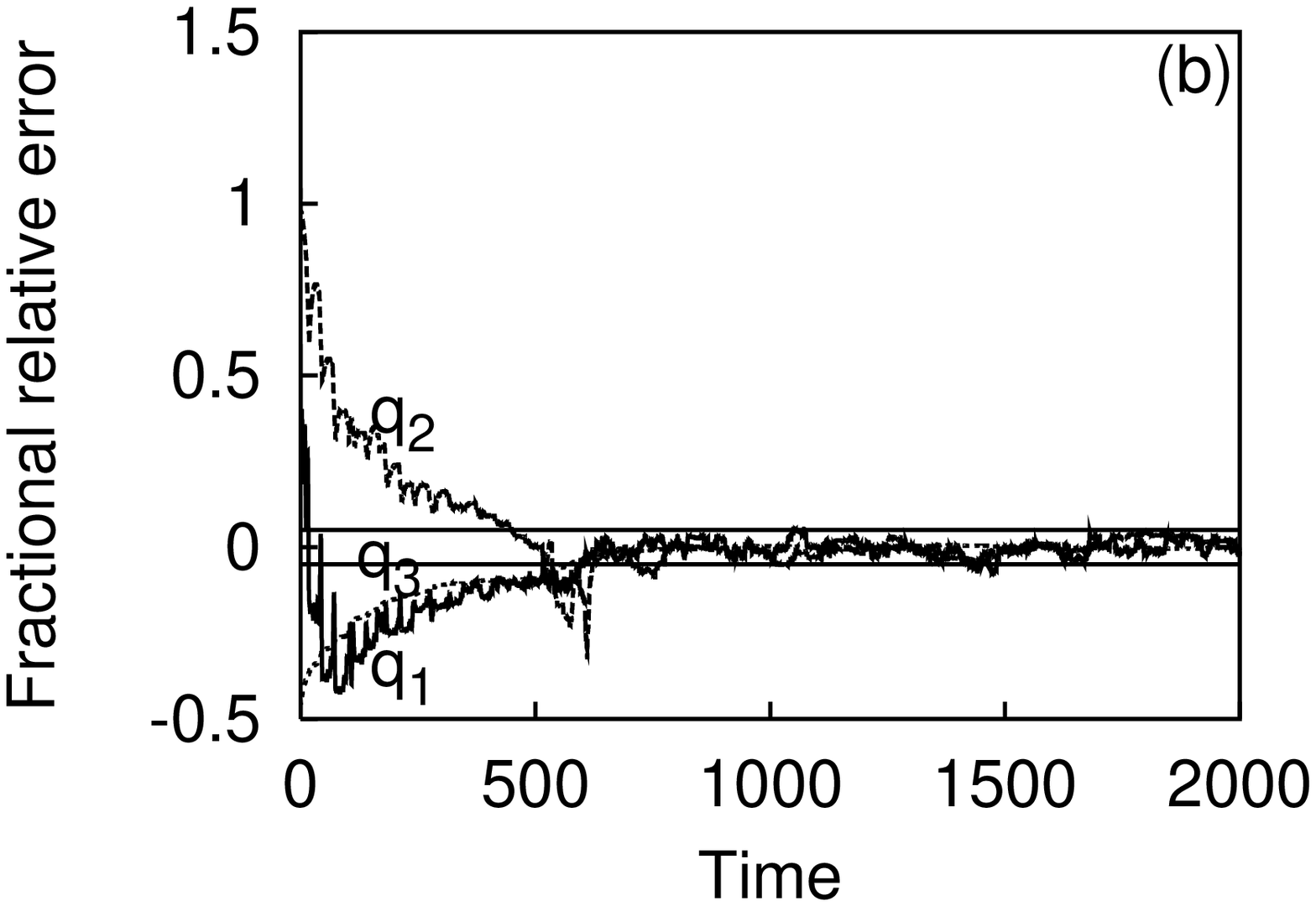,width=2.30in}}
\caption{Temporal evolution of the estimation errors for all the 
parameters of the Rossler system (9) for the case when (a) only
additive noise is present in the measured output $x_2$,
(b) additive noise is present and each of the parameters of the 
simulated
experimental system is increased by $10\%$ at $t=500$. The
parameter $q_3$ has been scaled down 50 times for greater clarity of
representation. 
The stability parameters are  $k=20$, $\epsilon_1=0.15$, 
$\epsilon_2=0.2$ and $\epsilon_3=2$. The straight lines
correspond to an error of $\pm 5\%$.}
\end{figure}
\newpage
\begin{figure}[htbp]
\centerline{\psfig{file=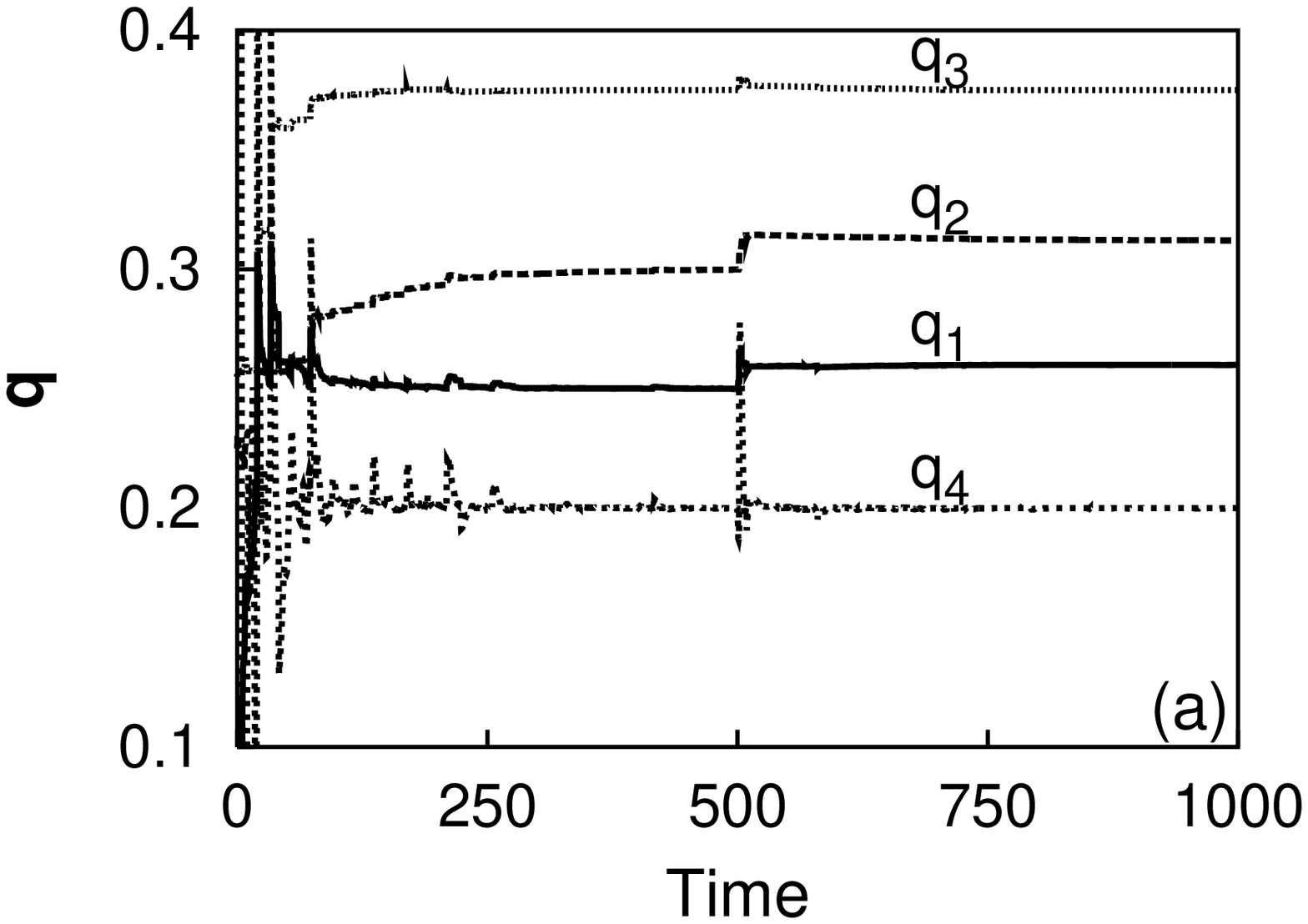,width=2.30in}}
\centerline{\psfig{file=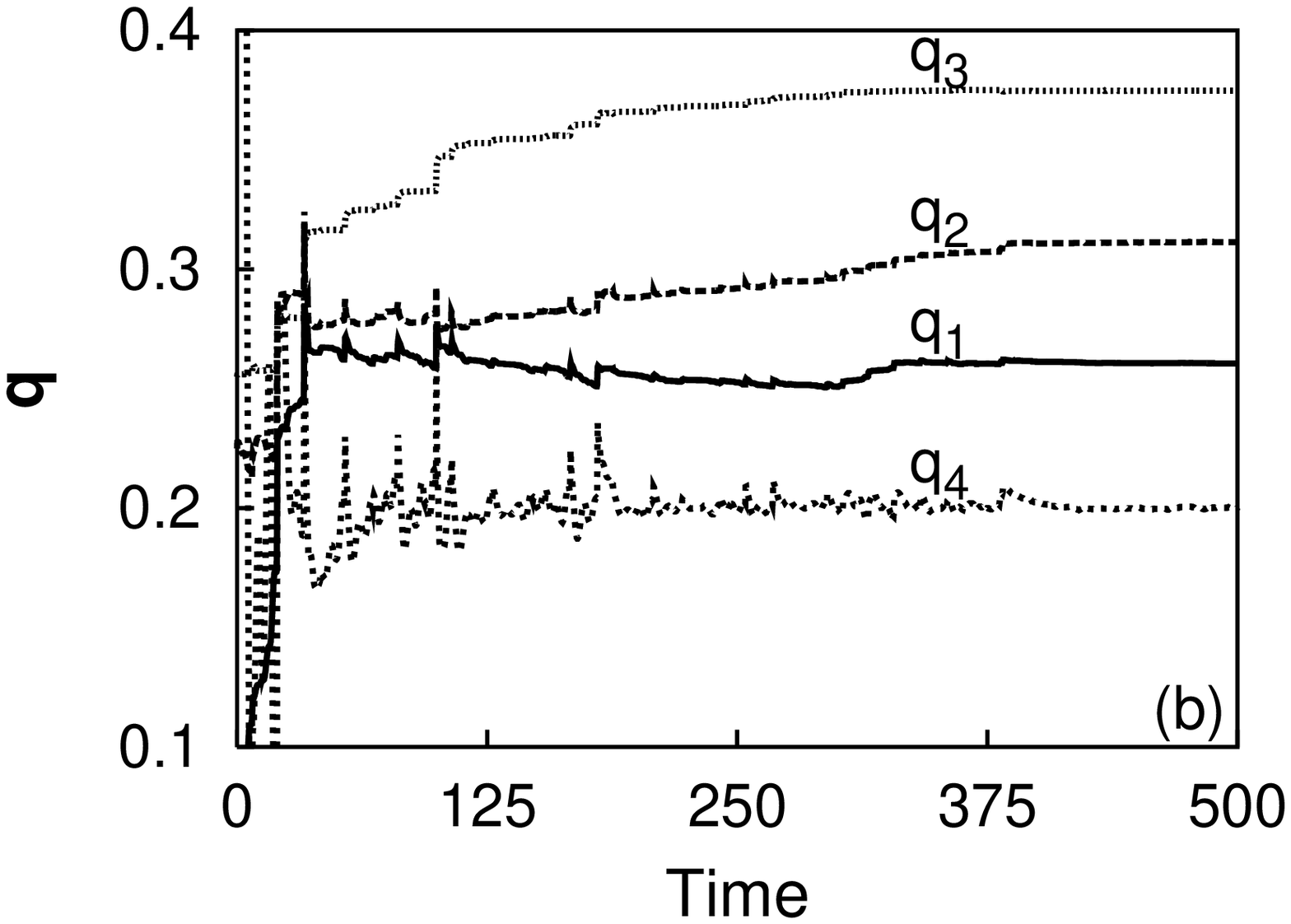,width=2.30in}}
\caption{Temporal evolution of the four parameters $q_1$, $q_2$,
$q_3$ and $q_4$ of the Rossler hyperchaos system (10) for the case
when (a) the parameters $p_1$ and $p_2$ are increased by $4\%$ at 
$t=500$, (b) additive noise is present in the 
measured output $x_2+x_4$, and the parameters $p_1$ and $p_2$ of the
simulated
experimental system are increased by $4\%$ at $t=300$.
For greater clarity of representation, the parameters have been
scaled in the following manner: $q_2$  and $q_3$  are scaled down
10 and $\frac{3}{4}$ times respectively, while the parameter $q_4$ 
is scaled up 4 times. The stability parameters are (a) $k=4$, 
$\epsilon_1=0.75$ and $\epsilon_2=\epsilon_3=\epsilon_4=0.005$; 
(b) $k=3.5$, $\epsilon_1=0.80$ 
and $\epsilon_2=\epsilon_3=\epsilon_4=0.002$.}
\end{figure}
\newpage
\begin{figure}[htbp]
\centerline{\psfig{file=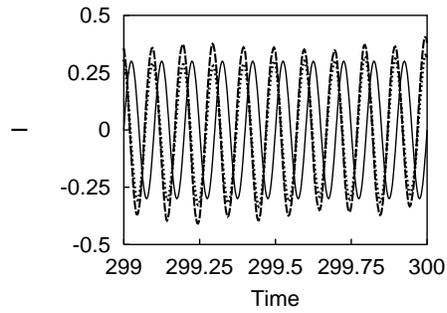,width=2.30in}}
\caption{Comparison of the message signal $I=\sin(20\pi t)$ and the
decoded signals obtained using the $x_2$ variable of Lorenz and 
Rossler models as the measured output. These are represented by the 
continuous, dashed and dotted lines respectively.  
Decoding results in phase shift as well as in a reduction
of the amplitude. The amplitude of the message
signal has been scaled for greater clarity. The stability parameters
are the same as that given in Fig. 1 and Fig. 2.}
\end{figure}
\end{document}